\documentclass[reprint, amsmath, amssymb, aps, pra]{revtex4-1}
\usepackage{physics}
\usepackage{xcolor}
\usepackage{graphicx}
\usepackage{dcolumn}
\usepackage{mathtools}
\usepackage{hyperref}

\usepackage{soul}

\begin{document}

\preprint{APS/123-QED}
\title{Selective tuning of Hilbert spaces in states encoded with spatial modes of light}
\author{Ali Anwar\textsuperscript{1}}\email{corresponding author, alianwarma@gmail.com}
\author{Nijil Lal\textsuperscript{1,2}}
\author{Shashi Prabhakar\textsuperscript{3}}
\author{R. P. Singh\textsuperscript{1}}

\affiliation{\textsuperscript{1}Quantum Science and Technology Laboratory (QST Lab), Physical Research Laboratory, Ahmedabad, India}
\affiliation{\textsuperscript{2}Indian Institute of Technology, Palaj, Gandhinagar, India}
\affiliation{\textsuperscript{3}Photonics Laboratory, Physics Unit, Tampere University, Tampere, FI-33720, Finland}

\date{\today}

\begin{abstract}
Spatial modes of light directly give the most easily accessible degree of freedom that span an infinite dimensional Hilbert space. The higher dimensional spatial mode entanglement realized using spontaneous parametric down conversion (SPDC) process is generally restricted to the subspace defined by a single spatial mode in pump. Access to other modal subspaces can be realized by pumping beams carrying several easily tunable transverse modes. As a proof of principle experiment, we generate twin-photon states in a nonlinear SPDC process with pump as a superposition of first order Laguerre-Gaussian (or Hermite-Gaussian) modes. We show that the generated states can be easily tuned between different subspaces by controlling the respective modal content in the pump superposition.
\end{abstract}

\pacs{Valid PACS appear here}
\maketitle

\section{\label{sec:level1}Introduction}
Photonic quantum information have made an explosive growth scientifically and technologically after the realization of various protocols utilizing various fundamental properties of photons. Experiments ranging from testing the fundamental principles of quantum mechanics to applications in quantum communication are easily realized by utilizing the correlations and entanglement observed between the twin photons (called \textit{signal} \& \textit{idler}) generated in a second order nonlinear optical process called spontaneous parametric down conversion (SPDC) \cite{boydbook}. In earlier days, more attention was given towards theoretical and experimental realization of two dimensional Hilbert space of photon polarization \cite{aspect, kwiat}. Two photon higher dimensional entanglement can be realized using various degrees of freedom such as time \cite{de2002creating}, frequency \cite{avenhaus} and space \cite{walborn, zhang}.

Entanglement in high dimensions are shown to have stronger Bell violations \cite{dada}, and can provide greater security \cite{cerf, bechmann} and higher information capacity \cite{bechmann, walborn, dixon}. Spatial modes of light are commonly used to represent higher dimensional quantum states. Entanglement of photon pairs in different spatial modes directly provides several choices of subspaces in a multidimensional Hilbert space. Most common among them are the Laguerre-Gaussian (LG) and Hermite-Gaussian (HG) modes. Light beams with LG mode have a cylindrical symmetry along propagation axis and have a doughnut-like intensity distribution. They have an associated azimuthal index that gives orbital angular momentum (OAM) of $l\hbar$ to each photon, where $l$ is the azimuthal index \cite{oneil_intrinsic_2002, allen, allen_orbital_1999}. In classical optics, such light beams are generally known as optical vortices. While, HG modes have a rectangular symmetry along the beam propagation \cite{dickey2018laser}. They each form a complete orthonormal basis with infinite dimensionality \cite{allen, andrews_babiker_2012, walborn2007transverse}. Access to higher dimensions in Hilbert space makes OAM of light a suitable candidate for realizing new types of secure quantum information schemes \cite{cerf, perumangatt2017quantum, zhangdecay}.

The theoretical and experimental aspects of correlations and entanglement of orbital angular momentum of photons are well established after initial experimental realizations of OAM entanglement of twin photons in a down conversion process \cite{franke, mair, mclaren}. A combination of a grating hologram, a single-mode fiber and a single photon detector acts as a mode projector for signal and idler photons in most of the quantum optics experiments involving spatial modes \cite{mair, jack}. Parametric down conversion experiments involving OAM generally use fundamental Gaussian pump and thereby measuring the correlations between the generated signal and idler photons having equal but opposite OAM values. Higher dimensional states of OAM are generally prepared by projecting signal and idler into appropriately designed holograms \cite{vaziri}. Dimension of the state generated by OAM entangled photons is determined by Schmidt decomposition \cite{law, torresqsb}. 
Interestingly, schemes were proposed for the preparation of photons in multidimensional vector states using optical vortex \textit{pancakes} made of Gaussian beam with distributions of nested azimuthal singularities in the pump \cite{molina}. As generation of paired photons is highly dependent on the pump beam and the nonlinear crystal, modal correlations can be manipulated by careful engineering of the pump \cite{torresqsb, torres} as well as the phase-matching parameters \cite{torresquasi}. Along with LG modes, correlations and entanglement of other spatial modes are further explored using different structured beams \cite{romero, kovlakov}. With an HG mode in the pump, two-photon maximally entangled states in HG basis are experimentally realized \cite{kovlakov} and interference of twin-photons in HG basis has been verified \cite{zhang-hom}.

A pump carrying a single spatial mode generates twin photons that are correlated in different spatial modes defined according to certain selection rules in SPDC \cite{walbornOAMconser, walbornHGconservation}. The spectrum constitutes a transverse subspace defined by a spatial mode in the pump. For a pump containing many spatial modes altogether, correlations between signal and idler will be present in the subspaces corresponding to each mode in the pump. So, multidimensional states with higher modal capacity can be easily generated by adding several modes to the pump. OAM correlations in multiple subspaces using phase-flipped Gaussian pump has been reported \cite{romero}. Recent experiments showed that maximally entangled qudits and ququarts can be generated by careful shaping of pump beam having additional spiral modes along with Gaussian using spatial light modulator \cite{liu2018coherent,kovlakov2018quantum}. 

In this work, we use interferometric technique \cite{chithra} to switch between different helical modes in the pump. A half-wave plate in the interferomter controls the tuning between modes in the pump. We give simple and generic theoretical description of tuning of pump modes in HG and LG basis using interferometry. The tuning leads to the corresponding switching among the subspaces that satisfies modal selection rules. We showed that the quantum spiral spectrum of twin photons contains multiple diagonals, where each diagonal corresponds to a spatial mode in the pump. Since LG modes can be expressed in terms of sums of HG modes \cite{oneil}, the same interferometric setup inherently gives a variable superposition of different HG modes in the pump. We also measured the two-photon spectrum in HG basis, which have states from two different HG subspaces.


The article goes as follows: Section \ref{theory} gives the detailed theoretical description of pump preparation, SPDC joint states in LG and HG bases and their projective measurements. The experimental details are given in section \ref{experiment}. The results of the experiments for selective tuning of states and further discussion on modal spectra in LG $\&$ HG bases are included in section \ref{results}. The presented work is concluded in section \ref{conclusion}.

\section{\label{theory}Theory}
A pump beam having a single spatial mode can be represented by the state $\ket{\psi_p}$. When a pump photon interacts in a nonlinear crystal, twin photons are generated with different spatial modes. The correlations between the modes is governed by certain spatial selection rules \cite{walbornOAMconser, walbornHGconservation}. Based on this, the joint signal-idler state can be represented as
\begin{equation}
    \ket{\psi}=\sum_j C_{ab}^{(j)}\ket{a_j}_s \ket{b_j}_i,
    \label{SPDC_single_sum}
\end{equation}
where $a_j$ and $b_j$ represent the $j$-th spatial modes of signal ($s$) and idler ($i$) photon respectively. $C_{ab}^{(j)}$ is the probability amplitude for occurrence of the state $\ket{a_j}_s\ket{b_j}_i$.

When pump contains superposition of many spatial modes, the state of the pump is written as $\ket{\psi_p}=\sum_{k}a_{k}\ket{f_k}$, then the corresponding SPDC state is given by
\begin{equation}
    \ket{\psi}=\sum_{k}G_k\sum_j C_{ab}^{(j)}\ket{a_j}_s \ket{b_j}_i,
    \label{SPDC_double_sum}
\end{equation}
where $G_k$ is the weight factor for each pump mode. For each mode $f_k$ in the pump, the $j$-summation in Eqn. \eqref{SPDC_double_sum} represents the biphoton eigenstate corresponding to $f_k$. In experiments, the probability amplitude corresponds to the coincidence counts obtained by projecting conjugate spatial modes in signal and idler. The coincidence counts representing $j$-th state is given by
\begin{equation}
    C_{ab}^{(j)}\propto {}^{}_{s}\langle a_j^{\text{(proj)}}\vert {}^{}_{i}\langle b_j^{\text{(proj)}}\vert \psi\rangle.
    \label{prob_ampl}
\end{equation}
Here, $a_j^{\text{(proj)}}$ and $b_j^{\text{(proj)}}$ represent spatial modes considered in projective measurement. In field representation, Eqn.\eqref{prob_ampl} is an overlap integral of the interacting pump ($E_p(\mathbf{k}_{p\perp}$)), signal ($E_s(\mathbf{k}_{s\perp}$)) and idler ($E_i(\mathbf{k}_{i\perp}$)), where $\mathbf{k}_{p\perp}$, $\mathbf{k}_{s\perp}$ and $\mathbf{k}_{i\perp}$ are respective transverse wave vectors that satisfy the momentum conservation
\begin{equation}
    \mathbf{k}_{p\perp}=\mathbf{k}_{s\perp}+\mathbf{k}_{i\perp}
    \label{mom_conservation}
\end{equation}
For a collinear phase-matched SPDC, the probability amplitude in Eqn.\eqref{prob_ampl} is rewritten as an overlap integral with the interacting fields
\begin{equation}
    C_{ab}^{(j)}=\int\int E_p(\mathbf{k}_{s\perp}+\mathbf{k}_{i\perp}) E_s^*(\mathbf{k}_{s\perp})E_i^*(\mathbf{k}_{i\perp})d\mathbf{k}_{s\perp}d\mathbf{k}_{i\perp}. \label{prob_ampl_num}
\end{equation}
The above integral depends on the momentum coordinates of signal and idler. In this paper, we consider the tunable superposition of two Laguerre-Gaussian (LG) modes as the pump for our down conversion process experiment.

\subsection{\label{pumpprep}Controlled generation of tunable superposition of  LG modes}
\begin{figure}[h]
    \centering
    \includegraphics[width=0.4\textwidth]{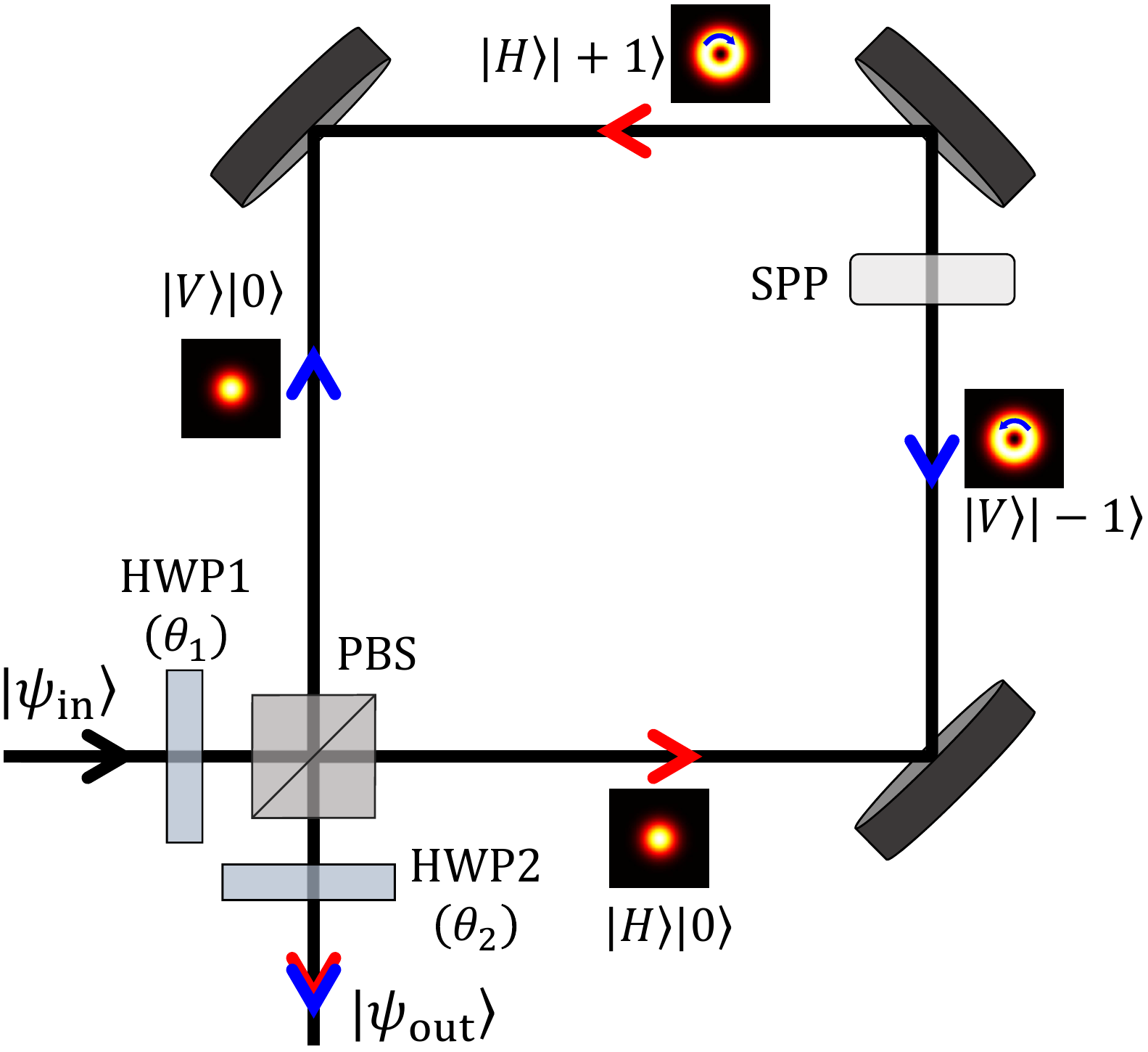}
    \caption[]{Experimental setup for the generation of variable superposition of optical vortices. HWP - Half wave plate; PBS - Polarizing beam splitter; SPP - Spiral phase plate. Counter clockwise propagating beam shown with red arrows and clockwise propagating beam is shown with blue arrows.} \label{nonsepexpt}
\end{figure}

For this purpose, we use the technique of polarization based Sagnac interferometer as shown in Figure \ref{nonsepexpt}. We start with a horizontally polarized Gaussian pump beam. The state of the laser beam before the first half-wave plate (HWP1) can be expressed as the tensor product of polarization and OAM bases,
\begin{equation}
    \ket{\psi_{\text{in}}}=\ket{H}\ket{0},
\end{equation}
where $\ket{H}$ represents the state of horizontally polarized input Gaussian beam and $\ket{0}$ denotes it's zero azimuthal index. Using Jones matrix notation, we can represent horizontally and vertically polarized light with the respective column vectors
\begin{equation}
    \vert H\rangle = \begin{pmatrix}1\\0\end{pmatrix} \quad;\quad \vert V\rangle = \begin{pmatrix}0\\1\end{pmatrix}, \label{jonesvect}
\end{equation} 
and the action of HWP1 whose fast axis is at an angle $\theta_1$ with respect to vertical axis is given by a $2\times2$ Jones matrix
\begin{equation}
    \hat{U}_{HWP}(\theta_1) = 
      \begin{pmatrix}
        \cos 2\theta_1 & \sin 2\theta_1 \\
        \sin 2\theta_1 & -\cos 2\theta_1 \\
      \end{pmatrix}.
    \label{jonesmat}
\end{equation}
After passing through HWP1, state of the beam becomes
\begin{equation}
    \hat{U}_{HWP}(\theta_1)\vert V\rangle\vert\ 0\rangle = \left(\sin 2\theta_1\vert H\rangle-\cos 2\theta_1\vert V\rangle\right)\vert 0\rangle .
    \label{out_hwp1}
\end{equation}

Now the beam is fed into a polarizing Sagnac interferometer where the orthogonally polarized beams ($\vert H\rangle$ \& $\vert V\rangle$) counter-propagate and combine at the output of the interferometer. A spiral phase plate (SPP) of winding order 1 converts the forward propagating Gaussian beam $\vert l=0\rangle$ to $\vert l=+1\rangle$ and back-propagating Gaussian beam $\vert l=0\rangle$ to $\vert l= -1\rangle$. So, the output state of generated beam after interferometer is
\begin{equation}
    \vert\Psi_{sag}\rangle = \sin 2\theta_1\vert H\rangle\vert +1\rangle+\cos 2\theta_1\vert V\rangle\vert-1\rangle . \label{out_sagnac}
\end{equation}
This light is then passed through the second half-wave plate (HWP2), whose fast axis is at an angle $\theta_2$ with respect to the vertical axis. Using similar calculations, we can find the state of output beam from Sagnac interferometer as
\begin{align}
    \vert\Psi_{out}\rangle =& \vert H\rangle\left(\sin 2\theta_1\cos 2\theta_2\vert +1\rangle+\cos 2\theta_1\sin 2\theta_2\vert -1\rangle\right) \nonumber \\
    +&\vert V\rangle\left(\sin 2\theta_1\sin 2\theta_2\vert +1\rangle-\cos 2\theta_1\cos 2\theta_2\vert -1\rangle\right). \label{out_hwp2_rearranged}
\end{align}
When the tuned pump in the state given by Eqn.\eqref{out_hwp2_rearranged} is incident on a non-linear crystal, it will down convert only the beam having polarization oriented along its optic axis. During the down conversion process, the crystal acts as a polarizer and down converts either $\vert H\rangle$ or $\vert V\rangle$ part of $\vert\Psi_{out}\rangle$. Considering the crystal to be aligned for down converting only $\vert H\rangle$ part, for a fixed HWP2 angle $\theta_2=\pi/8$, the pump state being down-converted is
\begin{equation}
    \vert\Psi_{p}\rangle = \frac{1}{\sqrt{2}}\left(\sin 2\theta_1\vert+1\rangle+\cos 2\theta_1\vert -1\rangle\right).
    \label{to_spdc_HWP2_fixed}
\end{equation}
The above equation represents a variable superposition of $\vert+1\rangle$ and $\vert-1\rangle$ modes. We can tune the OAM content in the beam between $\vert+1\rangle$ and $\vert-1\rangle$ by varying $\theta_1$. Then we measure the two photon spatial mode spectrum, in both Laguerre-Gaussian (LG) and Hermite-Gaussian (HG) bases.

\subsection{Projections in LG basis}
In field representation, Eqn. \eqref{to_spdc_HWP2_fixed}, can be rewritten in momentum coordinates $(\rho,\phi)$, as
\begin{align}
    E_p&= \frac{1}{\sqrt{2}}\left(\sin 2\theta_1LG_p^{1}(\rho,\phi)+\cos 2\theta_1LG_p^{-1}(\rho,\phi) \right),
    \label{Ep}
\end{align}
where $LG_p^l$ represents the Laguerre-Gaussian (LG) mode distribution of radial index $p=0$ and azhimuthal index $l$ given by \cite{allen}
\begin{align}
    LG_p^l(\rho,\phi)=&\sqrt{\frac{w^2p!}{2\pi (p+|l|)!}}\left(\frac{w\rho}{\sqrt{2}}\right)^{|l|}L_p^{|l|} \left(\frac{w^2\rho^2}{2}\right)\nonumber \\
    &\times\exp\left(-\frac{w^2}{4}\rho^2\right)\exp(il\phi),
\label{NOV}
\end{align}
where $L_p^{|l|}(.)$ is the associated Laguerre polynomial. In our experiment we have taken LG modes with radial index $p$=$0$ only. In LG eigenbasis, the spatial modes projected in signal and idler are respectively
\begin{equation}
    E_s=LG_0^{l_s}(\rho,\phi) \quad\text{and}\quad E_i=LG_0^{l_i}(\rho,\phi).
    \label{EsEi_LG}
\end{equation}
Substituting Eqn.\eqref{Ep} and Eqn.\eqref{EsEi_LG} in Eqn.\eqref{prob_ampl_num}, the three field overlap integral becomes
\begin{align}
    C_{l_s,l_i}=&\int_0^{2\pi}d\phi\int_0^\infty \rho d\rho \left(\sin 2\theta_1LG_0^{1}(\rho,\phi) \right .\nonumber \\
    &\left.+\cos 2\theta_1 LG_0^{-1}(\rho,\phi)\right) [LG_0^{l_s}(\rho,\phi)]^* [LG_0^{l_i}(\rho,\phi)]^*.
\end{align} 
The above integral simplifies to
\begin{align}
    C_{l_s,l_i}=&\sqrt{\frac{w_p^2}{\pi|l_s|!|l_i|!}}(\delta_{1,l_s+l_i}\sin 2\theta_1 +\delta_{-1,l_s+l_i}\cos 2\theta_1)\nonumber\\
    &\times
    \begin{dcases}
     \left(\frac{1}{3}\right)^{\alpha+1}\sqrt{\pi}\prod_{n=1}^{\alpha+\frac{1}{2}}(2n-1) & \text{if } 2\alpha \text{ is odd, or} \\[8pt]
     \left(\frac{2}{3}\right)^{\alpha+1}\alpha! & \text{otherwise,}
  \end{dcases}
  \label{LGequation1}
\end{align}

\begin{figure}[h]
    \begin{center}
        \includegraphics[width=0.4\textwidth]{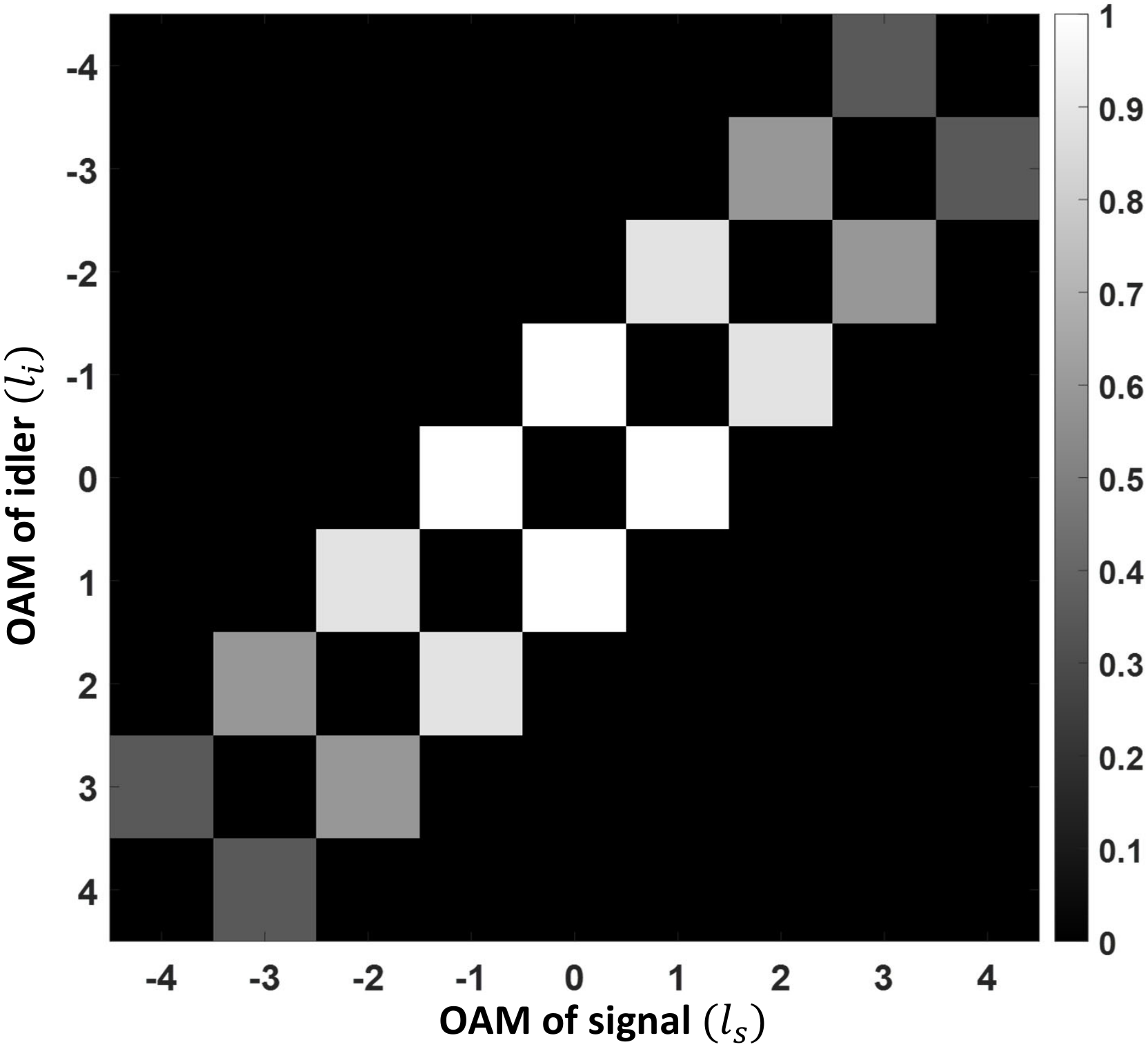}
        \caption[]{Theoretically estimated two-photon OAM spectrum ($C_{l_s,l_i}$) of SPDC photons in LG basis with equal superposition of +1 and $-1$ order optical vortex pump ($\theta_1=\pi/8$).}
        \label{OAM-theory}
    \end{center}
\end{figure}
where $\alpha$=$(1+|l_s|+|l_i|)/2$ and $\delta$ represents Kronecker delta function. For simplification, the pump and collection waists are taken to be same ($w_p$=$w_s$=$w_i$). Equation \eqref{LGequation1} is an exact expression in terms of $l_s$ and $l_i$. The twin photon's spectrum in LG basis based on Eqn.\eqref{LGequation1} is shown in Fig. \ref{OAM-theory}. The spectrum explains the independent OAM conservation relation for $l$=$+1$ and $l$=$-1$, where the sum of OAMs of pump, signal and idler photons equals zero. The decrease in the value of $C$ for higher $l$ values is due to the increase in overall size of the beams and results in less overall overlap. By varying the value of $\theta_1$, the spectrum in LG basis can be tuned easily and dynamically. The value of $\theta_1$ tunes the contribution to the above and below to diagonal series of overlap. The parameter $\theta_1$ is easy to manipulate in the experiment and a wide range of OAM spectrum can be obtained.

\subsection{Projections in HG basis}
A Hermite-Gaussian (HG) mode in momentum coordinates is written as
\begin{align}
    u_{m,n}(k_x,k_y)=&\frac{-i^{m+n}}{\sqrt{2^{m+n+1}\pi m!n!}}H_m\left(\frac{wk_x}{\sqrt{2}}\right) H_n\left(\frac{wk_y}{\sqrt{2}}\right) \nonumber \\
    &\cross \exp\left(-\frac{w^2}{4}(k_x^2+k_y^2)\right).
    \label{HG_momentum_coord}
\end{align}
To make calculation of overlap integral in Eqn.\eqref{prob_ampl_num} simpler, we consider the interacting fields in momentum coordinates. 
A LG mode from $\vert \Psi_p \rangle$ can be written in terms of sum of HG modes as in \cite{oneil}
\begin{align}
    LG_0^1&=\frac{1}{\sqrt{2}}(u_{1,0}(k_x,k_y)+iu_{0,1}(k_x,k_y)) \quad \text{and} \nonumber \\
    LG_0^{-1}&=\frac{1}{\sqrt{2}}(u_{1,0}(k_x,k_y)-iu_{0,1}(k_x,k_y)).
    \label{LG_HG}
\end{align}
Substituting Eqn.\eqref{LG_HG} in Eqn.\eqref{Ep}, we get
\begin{align}
    E_p(\mathbf{k}_{s\perp}+&\mathbf{k}_{i\perp})=\frac{1}{2}\left[(\sin 2\theta_1+\cos 2\theta_1)u_{1,0}(\mathbf{k}_{s\perp}+\mathbf{k}_{i\perp}) \right.\nonumber \\
    &+\left. i(\sin 2\theta_1-\cos 2\theta_1)u_{0,1}(\mathbf{k}_{s\perp}+\mathbf{k}_{i\perp})\right].
    \label{Ep_HG_final}
\end{align}
In an HG eigenbasis, the signal and idler modes are given as
\begin{align}
    E_s(\mathbf{k}_{s\perp})&= u_{m_s,n_s}(\mathbf{k}_{s\perp}), \nonumber \\
    E_i(\mathbf{k}_{i\perp})&= u_{m_i,n_i}(\mathbf{k}_{i\perp})
\end{align}
Then, the three beam overlap integral will be
\begin{align}
     C_{m_s,n_s}^{m_i,n_i}=&\int_{-\infty}^{\infty}\int_{-\infty}^{\infty}d\mathbf{k}_{s\perp}d\mathbf{k}_{i\perp}E_p(\mathbf{k}_{s\perp}+\mathbf{k}_{i\perp})\nonumber \\
    &\cross u_{m_s,n_s}^*(\mathbf{k}_{s\perp})u_{m_i,n_i}^*(\mathbf{k}_{i\perp})
    \label{HGoverlapint}
\end{align}
After evaluating the above integral (detailed calculation is shown in Appendix A), we get
\begin{align}
    C_{m_s,n_s}^{m_i,n_i}=\sqrt{\frac{\pi}{8}}\biggl[(\sin 2&\theta_1+\cos 2\theta_1)b(m_s,m_i,M-1) \nonumber \\
    \times b(&n_s,n_i,N)u_{M-1,N}(0,0) \nonumber \\ 
    & +i(\sin 2\theta_1-\cos 2\theta_1)b(m_s,m_i,M) \nonumber \\
    &\times b(n_s,n_i,N-1)u_{M,N-1}(0,0)\biggr]
    \label{C_HG_final}
\end{align}
where $M$=$m_s+m_i$, $N$=$n_s+n_i$ and
\begin{align}
    b(n,m,k)=&\sqrt{\frac{(n+m-k)!k!}{2^{n+m}n!m!}}\frac{1}{k!} \nonumber \\
    &\cross\dfrac{d^k}{dt^k}[(1-t)^n(1+t)^m]\vert_{t=0}. \label{bfunc}
\end{align}
Equation \eqref{C_HG_final}, together with Eqn. \eqref{bfunc}, forms the complete expression for projections of signal and idler modes in HG basis. Here, $u_{\gamma,\delta}$ is the HG mode \eqref{HG_momentum_coord} evaluated at $H_\gamma(0)$$\times $$H_\delta(0)$. For $\theta_1$=$\pi/8$, the second term in the Eqn.\eqref{C_HG_final} vanishes and the pump field will have only $HG_{1,0}$ term. When $\gamma$ is odd, the Hermite polynomial \cite{silvermanbook} $H_\gamma(0)=0$. So, for non-zero values of $u_{M-1,N}$, $M-1$ and $N$ must be even. Also, in the above expression, the coefficient $b$ will give non-zero values when the third independent variable in $b$ is a positive integer. Combining the two conditions for non-zero values of $C_{m_s,n_s}^{m_i,n_i}$, we can derive the selection rule for down converted fields as \cite{walbornHGconservation}
\begin{align}
    M=m_s+m_i\geq 1 \quad &; \quad \text{parity}(m_s+m_i)=odd \nonumber \\
    N=n_s+n_i\geq 0 \quad &; \quad \text{parity}(n_s+n_i)=even.
    \label{HG_selection_rule}
\end{align}
Evaluating for a particular value of index-set is easy. HG spectrum can also be engineered easily by tuning the HWP1 i.e. $\theta_1$.

\begin{figure}[h]
    \begin{center}
        \includegraphics[width=0.45\textwidth]{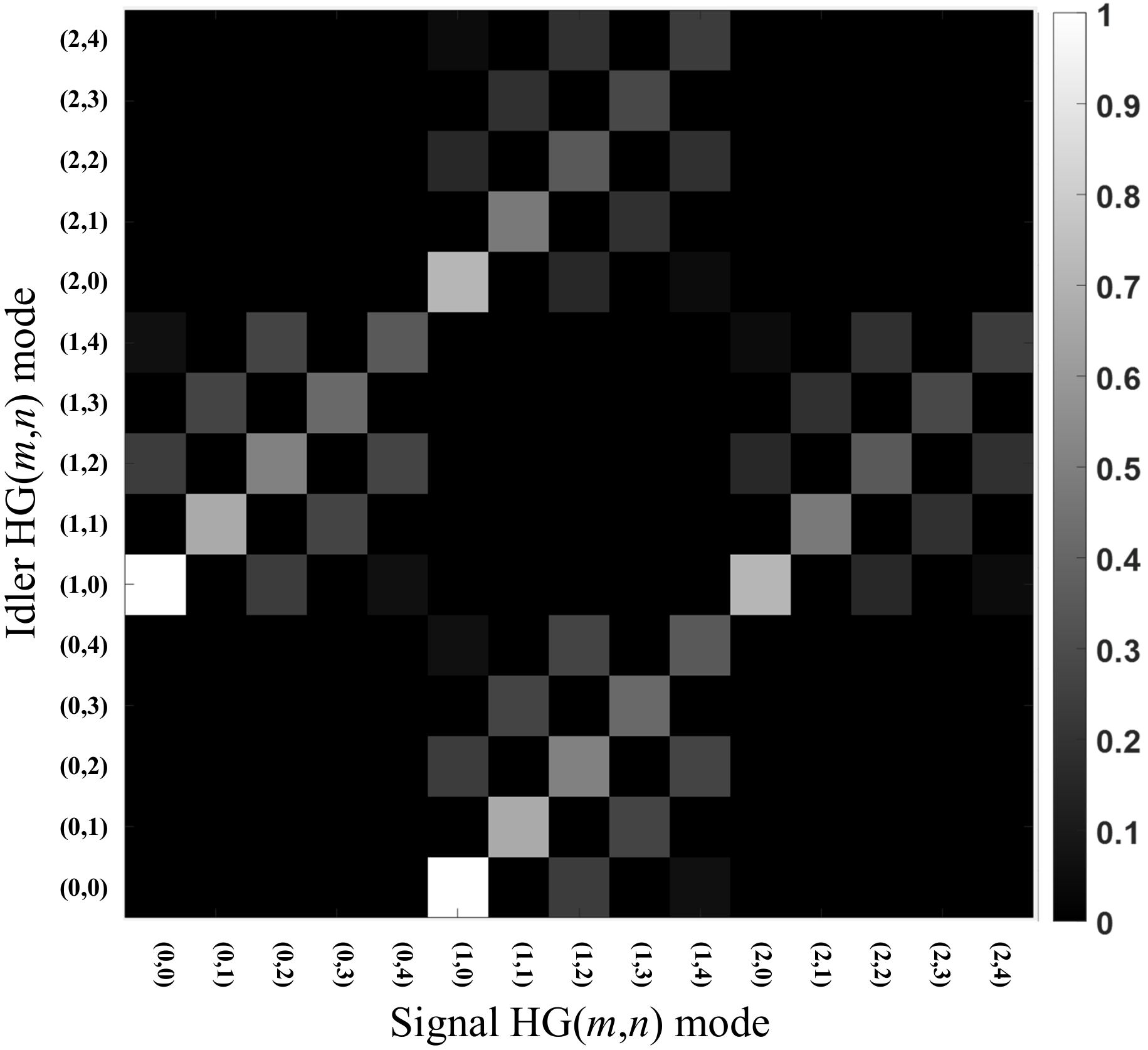}
        \caption[]{Theoretical twin photon OAM spectrum of SPDC photons in HG basis ($C_{m_s,n_s}^{m_i,n_i}$) with equal superposition of +1 and $-1$ order optical vortex pump ($\theta_1=\pi/8$).}
        \label{HG-theory}
    \end{center}
\end{figure}
Theoretically obtained spectrum in HG basis is shown in Fig. \ref{HG-theory}. This is the spectrum obtained for the HWP1 at $\theta_1$=$\pi/8$. The spectrum consists of two diagonals that corresponds to $HG_{1,0}$ part of the pump. From the spectrum, we can see that the values of $C$ are zero for both sum of indices, $m_s+m_i$ and $n_s+n_i$ having same parity (odd or even) or even and odd parities respectively.

\section{\label{experiment}Experiment}
\begin{figure}[t]
    \begin{center}
        \includegraphics[width=0.5\textwidth]{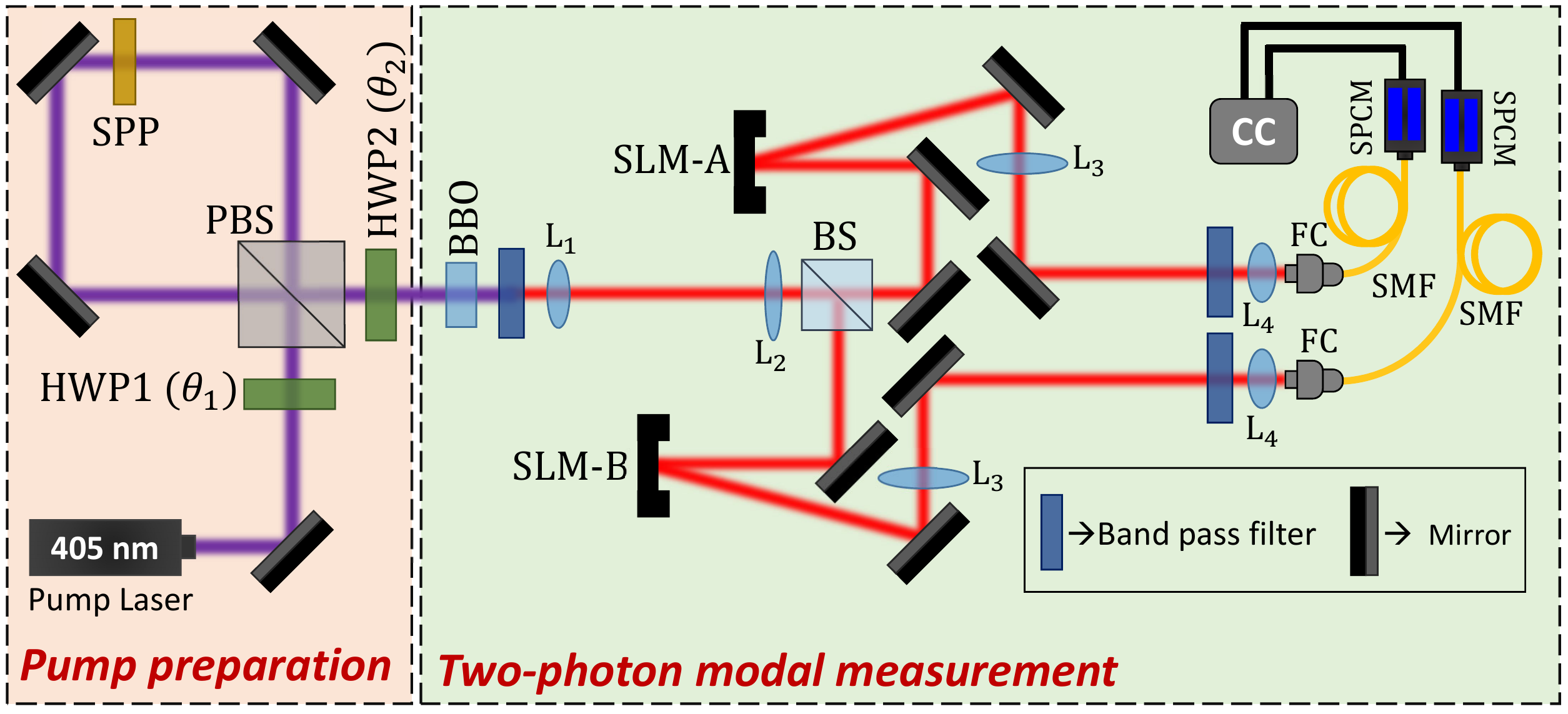}
        \caption[]{Experimental setup for the generation of required state and measurement of modal correlations in down-conversion process with pump as superposition of LG +1 and $-$1 modes. The setup for the pump preparation and the two-photon modal measurements are shown with different background colors.}
        \label{exptalsetup}
    \end{center}
\end{figure}

The experimental setup to measure spatial correlation between signal and idler photons for a pump of superposed vortices, is shown in Fig. \ref{exptalsetup}. The setup consists of a UV diode laser (Toptica iBeam smart) of wavelength 405 nm and power 250 mW with a spectral band-width of 2 nm. To generate required state, we set up a polarizing Sagnac interferometer as discussed in Section \ref{pumpprep}. The output beam prepared through the Sagnac interferometer pumps the non-linear crystal type-I $\beta$-Barium Borate (BBO) of thickness 5 mm. A band pass filter (BPF) of pass band 810$\pm$5 nm is used after the crystal to block pump beam and pass down converted photons. The down converted signal and idler photons of wavelength 810~nm each (degenerate pair) generated from the crystal, are imaged to spatial light modulators (SLM-A \& SLM-B) using lenses L$_1$ ($f$=100~mm) and L$_2$ ($f$=500~mm). SLMs are used to project the signal-idler pair to a particular LG/HG state. We select the first diffraction order of the output of each SLM so that the projected photons in the first order are Gaussian. This is achieved by imaging SLM plane to the fiber couplers (FC) in each arm using lenses L$_3$ ($f$=750~mm) and the aspheric lenses L$_4$ attached with the fiber coupler ($f$=2~mm).

The fiber couplers are attached to the single mode fibers (SMF) each having a numerical aperture of 0.13 and a mode field diameter of 5.0$\pm0.5~\mu$m, which in turn are connected to the single photon counting modules (SPCM, Excelitas SPCM-AQRH-16-FC). The SPCMs have a timing resolution of 350~ps with 25 dark counts per second. Two band pass filters of pass band 810$\pm$5~nm are kept very close to the fibre couplers to make sure that other unwanted wavelengths are properly filtered out. To measure the number of correlated photon pairs, the two detectors are connected to a coincidence counter (CC, IDQuantique ID800) having a time resolution of 81~ps. LabVIEW is used for the automation of projective measurements, in both LG and HG basis, by controlling the two modes on SLMs, performing the coincidence measurements from ID800, and recording the singles and coincidences in the text files.

\section{\label{results}Results and discussion}
From Eqn. \ref{to_spdc_HWP2_fixed}, we can see that the $\vert+1\rangle$ and $\vert-1\rangle$ state in pump varies periodically and this change is complementary to each other. This means that when $\vert+1\rangle$ OAM content is maximum, there is no contribution of $\vert-1\rangle$ OAM to the superposition, and vice versa. Considering only the non-zero probabilities in the LG spectrum, we can explicitly write the output OAM state of SPDC as
\begin{align}
    \ket{\psi}\propto &\cos 2\theta_1\biggl[C_{1,0}\ket{1}_s\ket{0}_i+C_{0,1}\ket{0}_s\ket{1}_i\nonumber \\
    &+C_{2,-1}\ket{2}_s\ket{-1}_i+C_{-1,2}\ket{-1}_s\ket{2}_i+...\biggr]\nonumber \\
    &+\sin 2\theta_1\biggl[C_{-1,0}\ket{-1}_s\ket{0}_i+C_{0,-1}\ket{0}_s\ket{-1}_i\nonumber \\
    &+C_{1,-2}\ket{1}_s\ket{-2}_i+C_{-2,1}\ket{-2}_s\ket{1}_i+...\biggr]
\end{align}
Here, the terms in the square brackets represent states in subspaces corresponding to $\vert+1\rangle$ and $\vert-1\rangle$ OAM in the pump, respectively.
\begin{figure}[t]
    \begin{center}
        \includegraphics[width=0.4\textwidth]{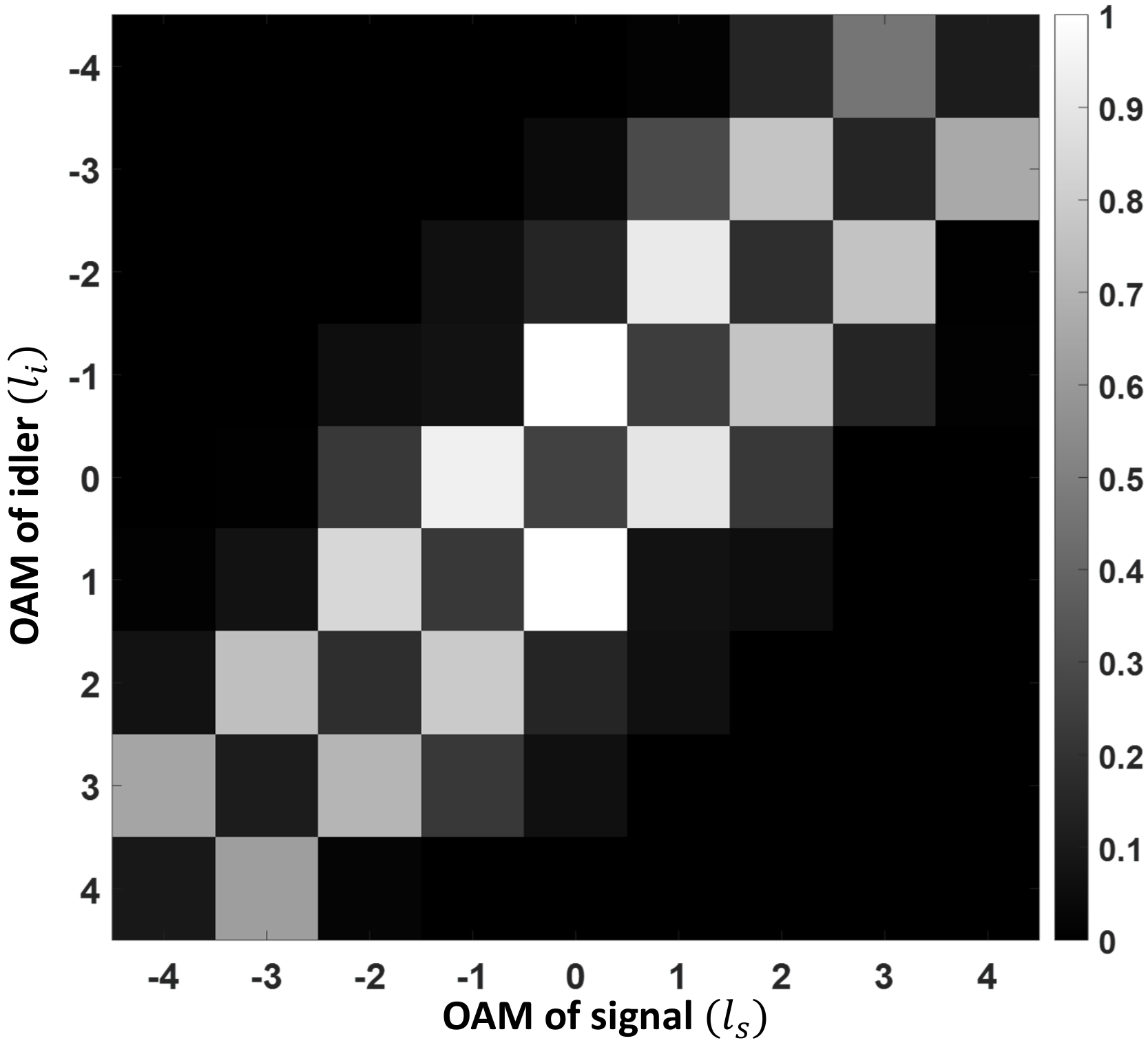}
        \caption[]{Experimentally measured two-photon OAM spectrum of SPDC photons in LG basis with equal superposition of +1 and $-1$ order optical vortex pump ($\theta_1=\pi/8$).}
        \label{OAM-exp}
    \end{center}
\end{figure}

To obtain the biphoton OAM spectrum experimentally, we recorded the coincidences of signal and idler photons coming in the first diffraction order of forked gratings on the two SLMs, coupled to single-mode fibres. For example, to obtain the probability for generating $\vert+1\rangle_a\vert 0\rangle_b$, we projected fork hologram of charge $-1$ in SLM1 and hologram of charge zero in SLM2. Due to the cancellation of equal but opposite azimuthal phases at the SLM, the amount of Gaussian light coupled into a single mode fibers in individual arms gives the singles counts, and the coincidence counts are measured with these. Figure \ref{OAM-exp} shows the experimentally measured OAM spectrum by performing projective measurements in signal and idler arms with OAM values ranging from $l=-4$ to 4. From theoretically estimated spiral spectrum in Fig.\ref{OAM-theory}, the Schmidt number is calculated to be $15.4\pm1.5$ and for experimental spectrum in Fig.\ref{OAM-exp}, it is $15.5\pm1.5$ \cite{straupe}. The values correspond to the effective number of accessible modes in the system, which here, is approximately the total number of non-zero diagonal elements in the spectra. For a spectra of states in a single subspace, the dimensionality will be almost half of that with two subspaces.

Considering the pump as a superposition of HG modes (as in Eqn.\eqref{Ep_HG_final}), and the selection rule for HG spectrum (Eqn.\eqref{HG_selection_rule}), biphoton state in HG basis can be written as
\begin{align}
    \ket{\psi}\propto &(\sin 2\theta_1+\cos 2\theta_1)\nonumber \\
    &\cross\biggl[C_{0,0}^{1,0}\ket{00}_s\ket{10}_i+C_{1,0}^{0,0}\ket{(10)}_s\ket{00}_i\nonumber \\
    &+C_{0,1}^{1,1}\ket{01}_s\ket{11}_i+C_{1,1}^{0,1}\ket{11}_s\ket{01}_i+...\biggr]\nonumber \\
    &+(\sin 2\theta_1-\cos 2\theta_1)\nonumber \\
    &\cross\biggl[C_{0,0}^{0,1}\ket{00}_s\ket{01}_i+C_{0,1}^{0,0}\ket{01}_s\ket{00}_i\nonumber \\
    &+C_{0,1}^{0,2}\ket{01}_s\ket{02}_i+C_{0,2}^{0,1}\ket{02}_s\ket{01}_i+...\biggr].
    \label{SPDC_HG}
\end{align}
For $\theta_1$=$\pi/8$, the pump field in terms of HG mode is obtained as
\begin{equation}
    E_p(\mathbf{k}_{s\perp}+\mathbf{k}_{i\perp})=\frac{u_{1,0}(\mathbf{k}_{s\perp}+\mathbf{k}_{i\perp})}{\sqrt{2}}.
    \label{Ep_HG_final_10}
\end{equation}
\begin{figure}[t]
\begin{center}
        \includegraphics[width=0.45\textwidth]{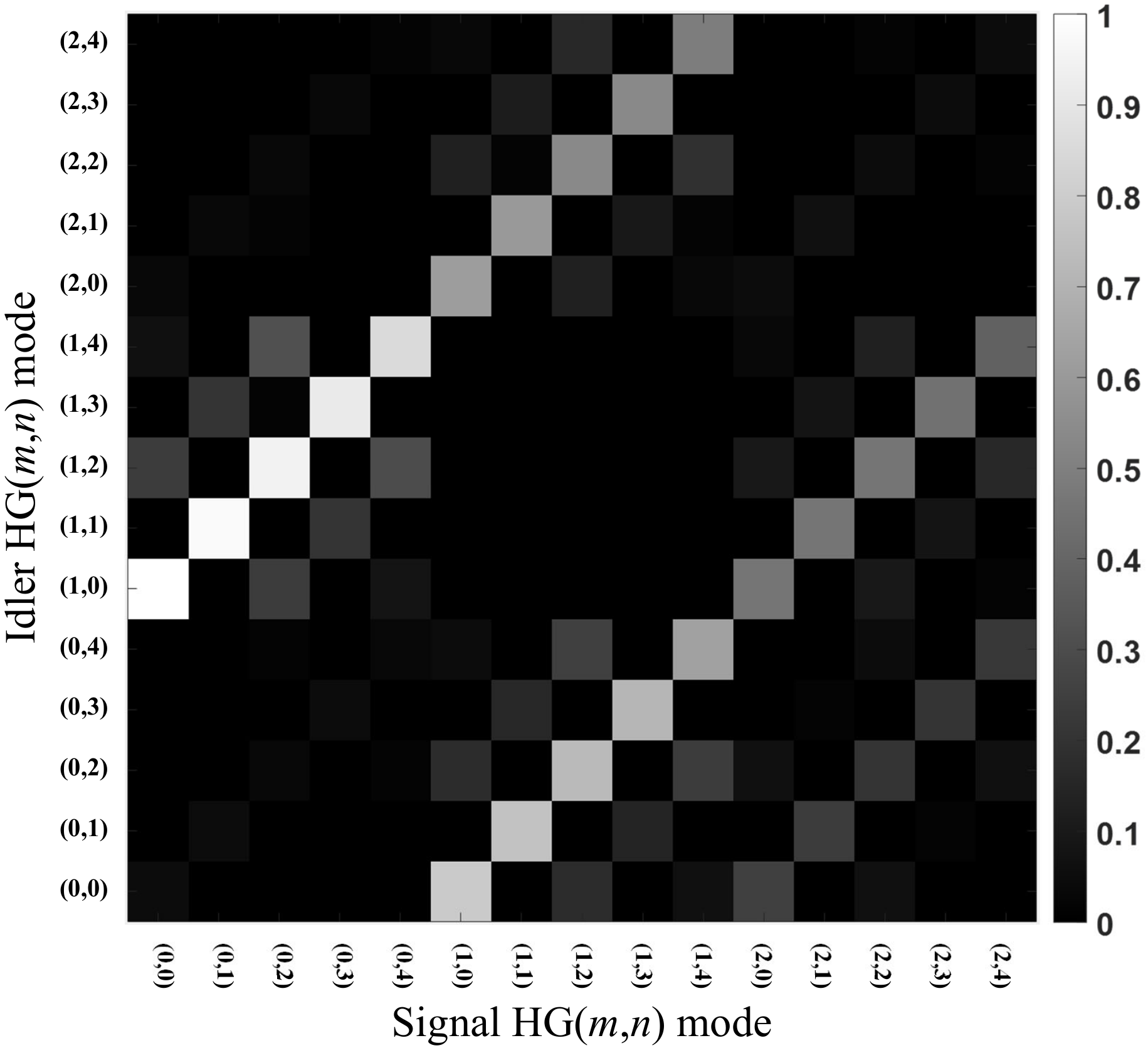}
        \caption[]{Experimentally measured two-photon spectrum of SPDC photons in HG basis with equal superposition of +1 and $-1$ order optical vortex pump ($\theta_1=\pi/8$).}
        \label{HG-exp}
    \end{center}
\end{figure}

For such a pump, the second term in Eqn.\eqref{SPDC_HG} vanishes and the contribution from the first term constitutes the spectrum. In experiment, we varied all the four indices of the two HG modes, two each for signal and idler ($m_s$, $m_i$, $n_s$ and $n_i$) and measured the coincidences. The measured HG spectrum is shown in Fig. \ref{HG-exp}. The observed mode spectrum  follows the mode selection rules given in Eqn.\eqref{HG_selection_rule}. Similar to the OAM spectrum, HG spectrum also contains two diagonals, which represent states corresponding to $HG_{1,0}$ pump mode. From the HG spectrum given in Fig.\ref{HG-theory}, the Schmidt number is estimated to be $32.7\pm7.8$. The experimental HG spectrum in Fig.\ref{HG-exp} gives a Schmidt number of $33.5\pm5.7$.

\begin{figure}[h]
    \begin{center}
        \includegraphics[width=0.45\textwidth]{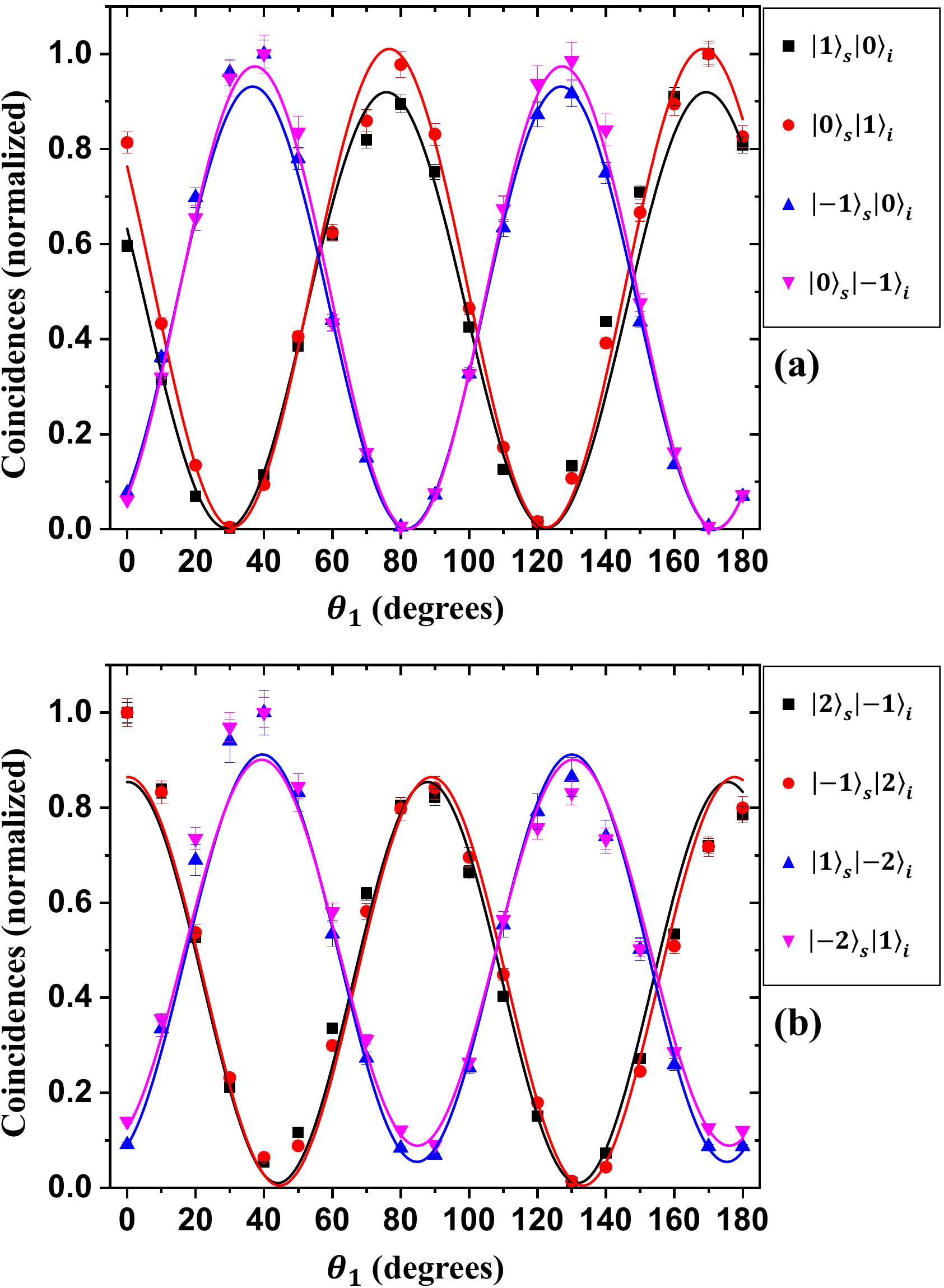}
        \caption{Plots of coincidence counts with respect to HWP1 angles for (a) $\ket{1}_s\ket{0}_i$, $\ket{0}_s\ket{1}_i$, $\ket{-1}_s\ket{0}_i$, $\ket{0}_s\ket{-1}_i$  and (b) $\ket{2}_s\ket{-1}_i$, $\ket{-1}_s\ket{2}_i$, $\ket{1}_s\ket{-2}_i$ and $\ket{-2}_s\ket{1}_i$ signal-idler OAM states.}
        \label{HWP1vsCC}
    \end{center}
\end{figure}

To verify the tuning of biphoton states further, we varied $\theta_1$ and recorded the coincidences in different signal-idler OAM bases. The coincidences for signal-idler OAM states $\ket{1}_s\ket{0}_i$, $\ket{0}_s\ket{1}_i$, $\ket{-1}_s\ket{0}_i$, $\ket{0}_s\ket{-1}_i$, $\ket{2}_s\ket{-1}_i$, $\ket{-1}_s\ket{2}_i$, $\ket{1}_s\ket{-2}_i$ and $\ket{-2}_s\ket{1}_i$ with respect to $\theta_1$ are plotted in Fig. \ref{HWP1vsCC}(a) \& (b). We observed a sinusoidal variation of coincidence counts with $\theta_1$. At the angles $\theta_1$ where $\ket{1}_s\ket{0}_i$ and $\ket{-1}_s\ket{0}_i$ coincidence curves intersect, the generated entangled state have equal contributions of $\ket{1}_s\ket{0}_i$, $\ket{0}_s\ket{1}_i$, $\ket{-1}_s\ket{0}_i$ and $\ket{0}_s\ket{-1}_i$. When $\theta_1$ is changed such that there is more $+1$ OAM and less $-1$ OAM in the pump, then the coincidences corresponding to $\ket{1}_s\ket{0}_i$ and $\ket{0}_s\ket{1}_i$, increase and that corresponding to $\ket{-1}_s\ket{0}_i$ and $\ket{0}_s\ket{-1}_i$ decrease.

Similar to the case of LG basis, we can see that the probabilities of occurrence of states corresponding to $HG_{10}$ and $HG_{01}$ pump varies from zero to maximum, when $\theta_1$ is tuned from $0^{\circ}$ to $180^{\circ}$. This results in the variation of effective number of accessible modes in HG basis. This is clear from Eqn.\eqref{SPDC_HG}. So, by controlling the amount of H- and V-polarized light in the interferometer, one can generate biphoton OAM state in particular selected bases in a controllable manner.

\section{\label{conclusion}Conclusion}
Spatial modes of light are one of the useful resources for encoding multidimensional quantum states to implement various quantum information protocols. The twin photon states in spatial modes generated via parametric down conversion process are generally restricted to a single modal subspace, due to the pump carrying a single spatial mode. We experimentally showed that the biphoton state can be spanned over multiple spatial eigenbases by adding several modes to the pump. We verified, both theoretically and experimentally, the mode spectrum in Laguerre-Gaussian basis and in Hermite-Gaussian basis. Both the modal spectra can be easily engineered using a half-wave plate. Further, the generated state can be switched among different OAM bases by controlling the amount of the individual OAM contribution in the pump superposition that corresponds to a particular basis.

We also verified the tuning of entangled states in modal basis with changing the angle of half wave plate. Calculated values of Schmidt numbers showed that the dimensionality of the tunable state with multiple subspaces is significantly higher than that with a single subspace. Also, the tuning of states in two different subspaces is complementary to each other, which directly comes from similar change to modal contribution in the pump. The presented results may find applications in the generation of novel higher dimensional OAM entangled states for quantum key distribution and communication.

\appendix \section{Calculation of $C_{m_s,n_s}^{m_i,n_i}$}\label{appendix}

To derive the analytical expression of $C_{m_s,n_s}^{m_i,n_i}$, we have adapted calculations from \cite{walbornHGconservation}. In the HG basis, the mode overlap integral is given by
\begin{align}
    C_{m_s,n_s}^{m_i,n_i}=&\int_{-\infty}^{\infty}\int_{-\infty}^{\infty}d\mathbf{k}_{s\perp}d\mathbf{k}_{i\perp}E_p(\mathbf{k}_{s\perp}+\mathbf{k}_{i\perp})\nonumber \\
    &\cross E_s^*(\mathbf{k}_{s\perp})E_i^*(\mathbf{k}_{i\perp})
    \label{HG_overlap_integral}
\end{align}
Here, $\mathbf{k}_{s\perp}$ and $\mathbf{k}_{i\perp}$ are the transverse wave vectors of signal and idler respectively. A HG mode in momentum coordinates is given in Eqn.\eqref{HG_momentum_coord} With Eqn.\eqref{Ep_HG_final} and \eqref{HG_momentum_coord}, Eqn.\eqref{HG_overlap_integral} becomes
\begin{align}
     C_{m_s,n_s}^{m_i,n_i}=&\frac{1}{2}\int_{-\infty}^{\infty}\int_{-\infty}^{\infty}d\mathbf{k}_{s\perp}d\mathbf{k}_{i\perp}\nonumber \\
     &\left[(\sin 2\theta_1+\cos 2\theta_1)U_{1,0}(\mathbf{k}_{s\perp}+\mathbf{k}_{i\perp}) \right.\nonumber \\
    &+\left. i(\sin 2\theta_1-\cos 2\theta_1)U_{0,1}(\mathbf{k}_{s\perp}+\mathbf{k}_{i\perp})\right]\nonumber \\
    &\cross u_{m_s,n_s}^*(\mathbf{k}_{s\perp})u_{m_i,n_i}^*(\mathbf{k}_{i\perp})
\end{align}
Here, $U_{m,n}(\mathbf{k})$ is equivalent to the expression \eqref{HG_momentum_coord} but characterized by the wavelength $\lambda_p$ and beam radius $w_p$ of the pump. To simplify the integral, we change the coordinates
\begin{equation}
    \mathbf{Q}=\mathbf{k}_{s\perp}+\mathbf{k}_{i\perp} \quad;\quad \mathbf{P}=\mathbf{k}_{s\perp}-\mathbf{k}_{i\perp},
\end{equation}
such that $d\mathbf{k}_{s\perp}d\mathbf{k}_{i\perp}$=$\frac{1}{2}d\mathbf{Q}d\mathbf{P}$. Then, we have
\begin{align}
     C_{m_s,n_s}^{m_i,n_i}=&\frac{1}{4}\int_{-\infty}^{\infty}\int_{-\infty}^{\infty}d\mathbf{Q}d\mathbf{P} \nonumber \\
     &\left[(\sin 2\theta_1+\cos 2\theta_1)U_{1,0}(\mathbf{Q}) \right.\nonumber \\
    &+\left. i(\sin 2\theta_1-\cos 2\theta_1)U_{0,1}(\mathbf{Q})\right]\nonumber \\
    &\cross u_{m_s,n_s}^*\left(\frac{\mathbf{Q}+\mathbf{P}}{2}\right)u_{m_i,n_i}^*\left(\frac{\mathbf{Q}+\mathbf{P}}{2}\right).
    \label{HG_overlap_integral2}
\end{align}
The interacting pump and SPDC fields in HG mode will have maximum overlap when $w_s$=$w_i$=$\sqrt{2}w_p$. For degenerate down conversion, $\lambda_s$=$ \lambda_i$=$2\lambda_p$. So, it can be easily shown that $u_{m,n}(\mathbf{k}/ \sqrt{2},\lambda_s,w_s)=u_{m,n}(\mathbf{k}/\sqrt{2},\lambda_i,w_i)=U_{m,n}(\mathbf{k},\lambda_p,w_p)$. Expanding $u_{m_s,n_s}^*$ and $u_{m_i,n_i}^*$ in Eqn.\eqref{HG_overlap_integral2} using Eqn.\eqref{HG_momentum_coord} and regrouping $x$ and $y$ terms, we get
\begin{align}
    u_{m_s,n_s}^*\left(\frac{\mathbf{Q}+\mathbf{P}}{2}\right)&u_{m_i,n_i}^*\left(\frac{\mathbf{Q}+\mathbf{P}}{2}\right) \nonumber \\
    =&U_{m_s,m_i}^*\left(\frac{Q_x+P_x}{\sqrt{2}},\frac{Q_x-P_x}{\sqrt{2}}\right)\nonumber \\
    &\cross U_{n_s,n_i}^*\left(\frac{Q_y+P_y}{\sqrt{2}},\frac{Q_y-P_y}{\sqrt{2}}\right).
    \label{HG_equvalence}
\end{align}
To simplify the integral further, we make use of diagonal Hermite-Gaussian (DHG) modes, defined as
\begin{align}
    DHG_{m,n}(\Tilde{k}_x,\Tilde{k}_y)=\sum_{\alpha=0}^{m+n}b(m,n,\alpha)u_{m,n}(k_x,k_y),
    \label{DHG_mode}
\end{align}
with $\Tilde{k}_x=(k_x+k_y)/\sqrt{2}$, $\Tilde{k}_y=(k_x-k_y)/\sqrt{2}$ and the coefficient $b(m,n,\alpha)$ defined as
\begin{equation}
    b(m,n,\alpha)=\sqrt{\frac{(m+n-\alpha)!\alpha!}{2^{m+n}m!n!}}\frac{1}{\alpha!}\dfrac{d^{\alpha}}{dt^{\alpha}}[(1-t)^m(1+t)^n]\vert_{t=0}.
\end{equation}
Substituting Eqn.\eqref{DHG_mode} on the RHS of Eqn.\eqref{HG_equvalence}, the product becomes
\begin{align}
    u_{m_s,n_s}^*\left(\frac{\mathbf{Q}+\mathbf{P}}{2}\right)&u_{m_i,n_i}^*\left(\frac{\mathbf{Q}+\mathbf{P}}{2}\right) \nonumber \\
    =&\sum_{\alpha=0}^Mb(m_s,m_i,\alpha)U_{M-\alpha,\alpha}^*(Q_x,P_x)\nonumber \\
    &\cross\sum_{\beta=0}^Nb(n_s,n_i,\beta)U_{N-\beta,\beta}^*(Q_y,P_y),
    \label{DHG_equvalence}
\end{align}
where $M=m_s+m_i$ and $N=n_s+n_i$. Substituting this Eqn.\eqref{DHG_equvalence} back in Eqn.\eqref{HG_overlap_integral2}, the coefficient $C$ becomes
\begin{align}
     C_{m_s,n_s}^{m_i,n_i}=&\frac{1}{4}\sum_{\alpha=0}^M\sum_{\beta=0}^Nb(m_s,m_i,\alpha)b(n_s,n_i,\beta)\nonumber \\
     &\cross\int_{-\infty}^{\infty}d\mathbf{Q}\left[(\sin 2\theta_1+\cos 2\theta_1)U_{1,0}(\mathbf{Q}) \right.\nonumber \\
    &+\left. i(\sin 2\theta_1-\cos 2\theta_1)U_{0,1}(\mathbf{Q})\right]\nonumber \\
    &\cross U_{M-\alpha,N-\beta}^*(\mathbf{Q})\int_{-\infty}^{\infty}d\mathbf{P}U_{\alpha,\beta}^*(\mathbf{P}).
    \label{HG_overlap_integral3}
\end{align}
The orthonormality condition of HG modes is
\begin{align}
    \int_{-\infty}^{\infty}d\mathbf{Q}U_{M-\alpha,N-\beta}^*&(\mathbf{Q})U_{m,n}(\mathbf{Q})=\delta_{M-\alpha,m}\delta_{N-\beta,n},
\end{align}
and by straightforward calculation of $\mathbf{P}$-integral in Eqn. \eqref{HG_overlap_integral3} using the series expansion of Hermite polynomial, the coefficient $C$ becomes

\begin{widetext}
\begin{align}
    C_{m_s,n_s}^{m_i,n_i}=\sqrt{\frac{\pi}{8}}\biggl[(\sin 2\theta_1+\cos 2\theta_1)&b(m_s,m_i,M-1)b(n_s,n_i,N)u_{M-1,N}(0,0) \nonumber \\
    &+i(\sin 2\theta_1-\cos 2\theta_1)b(m_s,m_i,M)b(n_s,n_i,N-1)u_{M,N-1}(0,0)\biggr]
\end{align}
\end{widetext}
where $b(.)$ is defined by Eqn. \eqref{bfunc}.


\end{document}